\documentclass[aps,prl]{revtex4}
\begin{document}

\title{Graviton propagator from background--independent quantum gravity\\[.5mm]} 
\author{\bf Carlo Rovelli}
\affiliation{Centre de Physique Th\'eorique de Luminy, Case 907, F-13288 Marseille, EU\\
Dipartimento di Fisica Universit\`a La Sapienza, P.A.$\,$Moro 2, I-00185 Roma, EU}
\date{\today}

\begin{abstract}

\noindent We study the graviton propagator in euclidean loop quantum gravity, using the spinfoam formalism.  We use boundary--amplitude and group--field--theory techniques, and compute one component of the propagator to first order, under a number of approximations, obtaining the correct spacetime dependence.  In the large distance limit, the only term of the vertex amplitude that contributes is the exponential of the Regge action: the other terms, that have raised doubts on the physical viability of the model, are suppressed by the phase of the vacuum state, which is determined by the extrinsic geometry of the boundary. 

\end{abstract}

\maketitle

\noindent An open problem in quantum gravity is to compute particle scattering amplitudes from the  background--independent theory and recover low--energy physics \cite{qg}. The difficulty is that general covariance makes conventional $n$-point functions ill--defined in the absence of a background.  A strategy for addressing this problem has been suggested in \cite{scattering}; the idea is to study the boundary amplitude, namely the functional integral over a finite spacetime region, seen as a function of the boundary value of the field \cite{oeckl}.  In conventional quantum field theory, this boundary amplitude is well--defined (see \cite{cr}) and codes the physical information of the theory; so does in quantum gravity, but in a fully background--independent manner \cite{cdort}. A generally covariant definition of $n$-point functions can then be based on the idea that the distance between physical points --arguments of the $n$-point function-- is determined by the state of the gravitational field on the boundary of the spacetime region considered.  In this paper we implement this strategy. 

We compute a first order term of the (connected) two-point function, starting from full non-perturbative quantum general relativity.  To this order, the computation produces a 4d version of the ``nutshell" 3d model studied in \cite{nutshell}.  Using a natural gaussian form of the vacuum state, peaked on the intrinsic \emph{as well as the} extrinsic geometry of the boundary, we derive an expression for a component of the graviton propagator. At low energy, this agrees with the conventional graviton propagator.  Other components and second order terms will be presented elsewhere \cite{leonardo}.  Our main motivation is to show that a technique for computing particle scattering amplitudes in background--independent theories can be developed. (The viability of the notion of particle in a finite region is discussed in \cite{daniele}.  For the general relativistic formulation of quantum mechanics underlying this calculation, see \cite{book}.  On the relation between graviton propagator and 3-geometries transition amplitudes in the conventional perturbative expansion, see \cite{Mattei:2005cm}.)   

We use some standard loop quantum gravity (LQG) results \cite{lqg,book}, as well as a specific  spinfoam model \cite{Perez,book}: the $GFT/B$ theory, in the terminology of \cite{book}, defined using group--field--theory methods. The results do not change using the $GFT/C$ model. These are tentative background--independent quantizations of euclidean general relativity. The first was introduced in \cite{DFKR} and is favored by some recent arguments \cite{reasons,Conrady:2005qu}. The second was introduced in \cite{PR} (see also \cite{PR2}) and has good finiteness properties \cite{finiteness}.   

The physical correctness of these theories has been questioned because in the low--energy limit  their interaction vertex (10$j$ symbol, or Barrett-Crane vertex amplitude) has been shown to include --beside the ``good" term approximating the exponential of the Einstein-Hilbert action  \cite{BarrettWilliams}-- also two ``bad"  terms: an exponential with opposite sign, giving the cosine of Regge action \cite{BarrettWilliams} (analogous to the cosine in the Ponzano--Regge model) and a dominant term that depends on the existence of degenerate four-simplices \cite{BCE,FL}. We show here that only the ``good" term contributes to the propagator. The others are suppressed by the rapidly oscillating phase in the vacuum state that peaks the state on its correct extrinsic geometry. Thus, the physical state selects the ``forward" propagating \cite{etera} component of the transition amplitude. This phenomenon was anticipated in \cite{colosi}. 

\vskip.3cm

Following \cite{scattering}, the 2-point function of quantum general relativity can be extracted from a nonperturbative formalism using the expression 
\begin{equation}
W^{a b c d}(x, y; q) = 
{\cal N} \sum_{ss'} \ W[s']\  \langle s' | h^{ab}(x)h^{cd}(y) |s\rangle 
\  \Psi_q[s].
 \label{simplypropaga2}
\end{equation}
Here $q$ is the intrinsic \emph{and} extrinsic (a coherent state depends on both classical position  \emph{and} momentum) geometry of a closed 3d surface, interpreted as the physical boundary of a 4d spacetime region, and $x$ and $y$ are points on this (metric) surface.  $W[s]$ is the boundary amplitude, that codes the quantum gravity dynamics, $\Psi_q[s]$ is a quantum state of the geometry peaked on $q$, the sum is over a basis $s$ of states of the 3-geometry, namely $s$-knot states, and $h^{ab}(x)$ is a linearized gravitational field operator acting on the state $ \Psi_q(s)$. We refer to  \cite{scattering} for a motivation and a discussion of this expression and to \cite{book} for notation and an introduction to the formalism and its ideas.  ${\cal N}$ is the normalization factor defined by $1 = {\cal N}  \sum_{s}  W[s]  \Psi_q(s)$.  Eq (\ref{simplypropaga2}) is well-defined if we have a dynamical model giving $W[s]$, a ``vacuum" state $\Psi_q[s]$ and a form for the operator $h^{ab}(x)$. 

We choose the boundary functional $W[s]$ defined by the group field theory $GFT/B$.
We recall the definition of $W[s]$,  referring to \cite{book} and \cite{Perez} for motivations and details.  The theory is defined for a field $\phi:(SO(4))^4\to R$ by an action of the form
\begin{equation} 
S[\phi] = S_{\rm kin}[\phi]  + \frac{\lambda}{5!}\  S_{\rm int}[\phi]. 
\end{equation} 
The field $\phi$ can be expanded in modes 
$\phi^{j_1\dots j_4}_{\alpha_1... \alpha_4\, i}$ (Eq.\,(9.71) of  \cite{book}). 
Indices $j_n,\ n=1,...,4$ label simple $SO(4)$ irreducible representations. Irreducible  representations of $SO(4)$ are labelled by a pair of spins $(j_+,j_-)$, corresponding to the split of $so(4)=su(2)\times su(2)$ into its self-dual and antiself-dual rotations; simple representations  are those where $j_+=j_-\equiv j$, and are therefore labelled by a single spin $j$.  The index $\alpha_n$ labels the components of vectors in the representation $j_n$. The index $i$ labels an orthonormal basis of intertwiners  (invariant vectors) on the tensor product of the four representations $j_n$.  We choose a basis in which one of the basis elements is the Barrett-Crane intertwiner $i_{BC}$ (Eq.\,(9.99) of  \cite{book}).    $SO(4)$-invariant observables of the theory are computed as the expectation values
\begin{equation}
W[s]  = \int \mathrm{D} \phi \ \ f_s(\phi) \  \  \mbox{e}^{-\int \phi^2 - \frac{\lambda}{5!} \int \phi^5}  , 
\label{ev}
\end{equation}
where $f_s(\phi)$ is the function of the field determined by a spinnetwork $s=(\Gamma,j_l,i_n)$. Recall that a spinnetwork is a graph $\Gamma$ formed by nodes $n$ connected by links $l$, colored with representations $j_l$ associated to the links and intertwiners $i_n$ associated to the nodes.    We indicate by $l_{nm}$ a link connecting the nodes $n$ and $m$ and by $j_{nm}\equiv j_{mn}$ the corresponding color.  The spinnetwork function is defined by 
\begin{equation}
f_s(\phi)= \sum_{\alpha_{nm}} \prod_n \phi^{j_{nm}}_{\alpha_{nm}i_n};
\end{equation}
here $n$ runs over the nodes and, for each $n$, the index $m$ runs over the four nodes that bound the four links $l_{nm}$ joining at $n$. Each index $\alpha_{nm} \equiv \alpha_{mn}$ appears exactly twice in the sum, and is thus contracted. The perturbative expansion of (\ref{ev}) in $\lambda$ leads to a sum over Feynman diagrams. The Feynman rules are given by the propagator
\begin{equation}
    \label{pph}
\mathcal{P}_{\alpha_ni}^{j_n}\,{}_{\alpha'_ni'}^{j'_n}= 
\delta_{i, i'}\  \sum_{\pi(n)}\
\prod_n\  \delta_{j_n, j'\!\!{}_{\pi(n)}}\  
\delta_{\alpha{}_{n}\alpha'\!\!{}_{\pi(n)}},
\end{equation}
where $\pi(n)$ are the permutations of the four numbers $n=1,2,3,4$;  
and the vertex amplitude
\begin{equation}
    \label{vertex}
\mathcal{V}^{\alpha_{nm}i_n}_{j_{nm}}= \Big(\prod_{n}
\delta_{i_{n}i_{\rm BC}}\Big)
\Big(\prod_{n<m}
\delta_{\alpha_{nm}\alpha_{mn}}\Big)
\ {\mathcal{B}}(j_{nm}),
\end{equation}
where the index $n=1,...,5 $ labels the five legs of the five-valent vertex; while the index $m\ne n$ labels the four indices on each leg. $ {\mathcal{B}}(j_{nm})$ is the 10$j$ symbol, or Barrett--Crane vertex amplitude \cite{Perez}, whose large $j$ expansion is discussed below.  The sums over permutations in the propagator give rises to a number of terms. Each of these can be interpreted as a spinfoam $\sigma$, by identifying closed sequences of contracted deltas as faces. Hence the amplitude (\ref{ev}) can be written as a sum of amplitudes of spinfoams bounded by a given spinnetwork 
$
W[s]  = \sum_{\partial\sigma=s}\ A[\sigma],
$
an expression that is naturally interpreted (and can also be derived) as a sum over discretized 4-geometries bounded by a given discretized 3-geometry, namely as a definition of the Misner-Hawking sum-over-geometries formulation of quantum gravity, where the amplitude of a 3-geometry ${}^3\!g$ is given by the Feynman integral over 4-geometries
$
W[{}^3\!g]  = \int_{\partial g={}^3\!g} Dg \  e^{iS[g]}.
$
The amplitude $A[\sigma]$ of a spinfoam $\sigma$ with $n$ vertices is proportional to $\lambda^n$.   Below we consider the first order of the expansion in $\lambda$.  
Since each face $f$ gives then a contribution $\dim j_f$ (by contracting the deltas), 
in the large $j$ limit the dominant term is the one with the largest number of faces.  
To first order in $\lambda$, we have only one possibility: one vertex and a five-valent boundary spinnetwork.   The dominant contribution for large $j$ is obtained by choosing the spinfoam $\sigma$ and the spinnetwork $s=\partial\sigma$ defined by the dual of a four-simplex and its boundary. The spinnetwork  $s$  has five nodes, connected by 10 links as follows
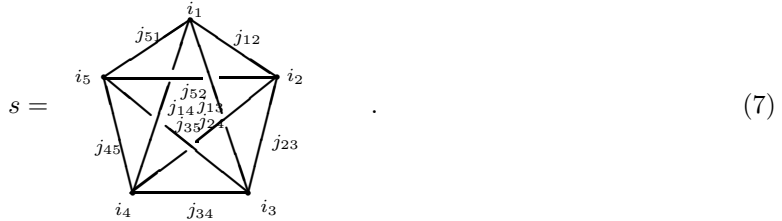
\begin{equation}
s = \hspace{-4em}\setlength{\unitlength}{0.0005in} 
\begin{picture}(5198,1100)(5000,-4330)
\thicklines
\put(8101,-5161){\circle*{68}}
\put(8401,-3961){\circle*{68}}
\put(6601,-3961){\circle*{68}}
\put(7501,-3361){\circle*{68}}
\put(6901,-5161){\circle*{68}}
\put(8101,-5161){\line(1,4){300}}
\put(7501,-3361){\line( 3,-2){900}}
\put(6601,-3961){\line( 3, 2){900}}
\put(8101,-5161){\line(-1, 0){1200}}
\put(6901,-5161){\line(-1, 4){300}}
\put(7876,-4479){\line( 1,-3){225.800}}
\put(7089,-4351){\line(-6, 5){491.312}}
\put(7801,-3961){\line( 1, 0){600}}
\put(6901,-5161){\line( 1, 3){383.100}}
\put(7321,-3871){\line( 1, 3){173.100}}
\put(7501,-3354){\line( 1,-3){325.500}}
\put(7456,-4726){\line(-4,-3){581.920}}
\put(8394,-3961){\line(-5,-4){828.902}}
\put(7569,-4629){\line( 0,-1){  7}}
\put(6601,-3969){\line( 1, 0){1020}}
\put(8109,-5161){\line(-5, 4){867.073}}
\put(7456,-3264){\makebox(0,0)[lb]{\smash{${}_{i_1}$}}}
\put(8499,-3961){\makebox(0,0)[lb]{\smash{${}_{i_2}$}}}
\put(8229,-5326){\makebox(0,0)[lb]{\smash{${}_{i_3}$}}}
\put(6300,-3969){\makebox(0,0)[lb]{\smash{${}_{i_5}$}}}
\put(6714,-5349){\makebox(0,0)[lb]{\smash{${}_{i_4}$}}}
\put(7951,-3551){\makebox(0,0)[lb]{\smash{${}_{j_{12}}$}}}
\put(8349,-4644){\makebox(0,0)[lb]{\smash{${}_{j_{23}}$}}}
\put(7464,-5356){\makebox(0,0)[lb]{\smash{${}_{j_{34}}$}}}
\put(6504,-4696){\makebox(0,0)[lb]{\smash{${}_{j_{45}}$}}}
\put(6929,-3521){\makebox(0,0)[lb]{\smash{${}_{j_{51}}$}}}
\put(7569,-4261){\makebox(0,0)[lb]{\smash{${}_{j_{13}}$}}}
\put(7339,-4462){\makebox(0,0)[lb]{\smash{${}_{j_{35}}$}}}
\put(7266,-4284){\makebox(0,0)[lb]{\smash{${}_{j_{14}}$}}}
\put(7576,-4423){\makebox(0,0)[lb]{\smash{${}_{j_{24}}$}}}
\put(7397,-4110){\makebox(0,0)[lb]{\smash{${}_{j_{52}}$}}}
\end{picture}\!\!\!\!\!\!\!\!\!\!\!\!\!\!\!\!\!\!.
\label{fsimpic}
\end{equation}\vskip1.1cm\noindent
The boundary function $f_s(\phi)$ determined by this spinnetwork is  
$f_s(\phi) = \sum_{\alpha_{nm}}\prod_{n=1,5} 
\phi^{\alpha_{nm}  i_n}_{j_{nm}}$. This is a an observable in the group field theory. The dominant term of its expectation value (\ref{ev}) is
\begin{equation}
W[s]  =  \frac{\lambda}{5!} \left( \prod_{n}   \langle i_n  |i_{\rm BC}\rangle   \right) 
              \left(\prod_{n < m} {\rm dim}({j_{nm}})\right) \, 
            \mathcal{B}(j_{nm}).
 \label{simply1}
  \end{equation}    

The second ingredient we need is a boundary state $\Psi_q[s]$. To identify it, we need the geometrical interpretation of the boundary spinnetwork $s$. To this aim, we use the fact that the spinfoam model can be obtained from a discretization of general relativity on a triangulated  spacetime.  Introduce 4d coordinates $x^\mu$ and represent the gravitational field by means of the one-form tetrad field $e^I(x) = e^I_\mu(x)dx^\mu$ (related to Einstein's metric by $g_{\mu\nu}(x)=e^I_\mu(x)e_I{}_\mu(x)$).   Assuming that the triangulation is fine enough for this field to be approximately constant on a tetrahedron, with constant value $e^I_\mu$, associate the 4d vector $e_{\rm s}^I=e^I_\mu \Delta x^\mu_{\rm s}$ to the segment ${\rm s}$ of the triangulation, where $\Delta x^\mu_{\rm s}$ is the coordinate difference between the initial and final extremes of ${\rm s}$. To each triangle $t$ of the triangulation, associate the bivector (the object with two antisymmetric indices)
\begin{equation}
B^{IJ}_{t} = e_{\rm s}^I e_{\rm s'}^J - e_{\rm s}^J e_{\rm s'}^I ,
\label{B}
\end{equation}
where $\rm s$ and $\rm s'$ are two sides of the triangle. (As far as orientation is kept consistent, the choice of the sides does not matter). $B^{IJ}_{t} $ is a discretization of the Plebanski two-form $B^{IJ} = e^I\wedge e^J$.  The quantum theory is then formally obtained by choosing the quantities $B^{IJ}_{t}$ as basic variables, and identifying them with $SO(4)$ generators $J^{IJ}_{t}$ associated to each triangle of the triangulation, or, equivalently, to each face of the corresponding dual spinfoam. The form (\ref{B}) implies that $\epsilon_{IJKL}B^{IJ}_{t}B^{KL}_{t'} =0$ any time $t=t'$ or $t$ and $t'$ share an edge. Accordingly, the pseudo--scalar Casimir $\tilde C=\epsilon_{IJKL}J^{IJ}_{t} J^{KL}_{t}=0$ is required to vanish. This determines the restriction to the simple representations, which are the ones for which $\tilde C=0$. The scalar Casimir $C=\frac{1}{2}J^{IJ}_{t} J_{t}{}_{IJ}=\frac{1}{2}B^{IJ}_{t} B_{t}{}_{IJ}$, on the other hand, is easily recognized, using again (\ref{B}), as the square of the \emph{area} $A_{t}$ of the triangle $t$.   For simple representations, the value of $C$ is $j(j+1)$. The quantization of the physical area, with eigenvalues proportional to $\sqrt{j(j+1)}$ is a key result of LQG \cite{lqg}, reappearing here in the context of the spinfoam models. The LQG result assures us that we can interpret it as a physical quantization and not an artifact of the discretization and fixes the proportionality constant: $A_j=
8\pi\hbar G \sqrt{j(j+1)}$. The geometrical interpretation of the intertwiners can be obtained in a similar way and will be discusses elsewhere. 

Let now $q$ be the intrinsic and extrinsic geometry of the boundary $\Sigma_q$ of a 4d (metric) sphere.  We want to construct the state $\Psi_q[s]$. (On the vacuum states in LQG, see \cite{vuoto1,vuoto2,vuoto3,vuoto4,vuoto5}.)  Below we shall only need the value of $\Psi_q[s]$ for the spinnetworks defined on a graphs $\Gamma$, dual to 3d triangulations $\Delta$.  We identify each such $\Delta$ with a fixed triangulation of $\Sigma_q$. 
We assume here for simplicity that  $\Psi_q[s]$ is non vanishing only for $s$ whose intertwiners are $i_{\rm BC}$ intertwiners.  This choice preserves the symmetry of the sphere.  (There is another simple symmetric choice,
used in the first version of this draft, which is to choose $\Psi_q[s]$ independent from the intertwiners.  This possibility will be explored elsewhere.)  The areas $A_{nm}$ of the triangles $t_{nm}$ of $\Delta$ determine background values $j^{(0)}_{nm}$ of the spins, via $A^2_{nm} = (8\pi\hbar G)^2\ j^{(0)}_{nm} (j^{(0)}_{nm} +1)$.  We want a state $\Psi_q[s] = \Psi_q(\Gamma, j_{nm})$ peaked on these background values.   The simplest possibility is to choose a gaussian peaked on these values. However, this leaves the possibility open of having a phase
\begin{equation}
\Psi_q[s] = 
 \exp\left\{-\frac{1}{2j_L}\alpha_{(nm)(pq)}(j_{nm}- j^{(0)}_{nm})(j_{pq}- j^{(0)}_{pq})
+i \sum_{n<m}\Phi^{(0)}_{nm} j_{nm}\right\}
\label{vuoto}
\end{equation} 
where $(n,m)$ runs on links of $s$, $\alpha_{(nm)(pq)}$ is a numerical matrix that we will fix later on and the sum $\sum_{n<m,p<q}$ is understood. 
 The phase factor in this state is important. As we know from elementary quantum mechanics, it determines where the state is peaked in the variables conjugate to the spins $j_{mn}$. Recall the form of the Regge action $S_{\rm Regge}=\sum_{n<m}\Phi_{nm}(j_{mn}) j_{nm}$, where $\Phi_{nm}(j_{mn}) $ are dihedral angles at the triangles, which are function of the areas themselves and that 
$\partial S_{\rm Regge}/\partial j_{nm}=\Phi_{nm}.$  It is then easy to see that these dihedral angles are the variables conjugate to the spins. Notice also that they code the extrinsic geometry of the boundary surface, and in GR the extrinsic curvature is the variable conjugate to the 3-metric.  The value of  $\Psi_q[s]$ on the five--valent spinnetwork (\ref{fsimpic}) can be determined by triangulating $\Sigma_q$ with the 3d triangulation formed by the boundary of a \emph{regular} four--simplex of side $L$.  The area of the triangles is $A_L=\sqrt{3}L^2/4$.  Then $j^{(0)}_{nm}=j_L$ where $8\pi \hbar G \sqrt{j_L(j_L+1)}=A_L$.  In the large $L$ limit we take $j_L=8\pi\hbar GA_L$.  The dihedral angles $\Phi_{nm}^{(0)}=\Phi$ of a regular tetrahedron are given by $\cos(\Phi) = -1/4$. 
 The $1/j_L$ dependence of $\alpha$ (absent in the first version of this paper) ensures that the relative uncertainties of areas and angles --that is: intrinsic and extrinsic geometry of the boundary-- become small in the large distance limit. Its need has been pointed out by Simone Speziale in the 3d context \cite{Simone} and by John Baez in the 4d case, following numerical investigation by Dan Christensen and Greg Egan, that have shown that in the absence of this dependence the width of the gaussian is not sufficient for the approximation taken above to hold \cite{JDC}.
To respect the symmetry of the symmetry of the sphere, the covariance matrix $\alpha_{(nm)(pq)}$ of the gaussian can depend only on three numbers 
\begin{equation}
\alpha_{(nm)(pq)}= \alpha_1\ a_{(nm)(pq)}+ \alpha_2\ \delta_{(nm)(pq)}  +  \alpha_3\ b_{(nm)(pq)} 
\label{simmetry}
\end{equation}
where $\delta_{(nm)(pq)}=1$ if $(nm)=(pq)$, 
$a_{(nm)(pq)}=1$ if just two indices are the same, 
and $b_{(nm)(pq)}=1$ if all four indices are different,  and in all other cases these quantities vanish. 
(We will use this notation, namely $\alpha_{(12)(13)}=\alpha_1, \ \alpha_{(12)(12)}=\alpha_2, \ \alpha_{(12)(34)}=\alpha_3 $ repeatedly.) The state (\ref{vuoto}) is thus completely determined up to the three numbers $\alpha_1,\alpha_2,\alpha_3$. This amounts to select a vacuum state which is a coherent state peaked both on the background values of the spins (the intrinsic geometry of the boundary surface), \emph{and} on the background values of the angles (the intrinsic geometry of the boundary surface). See \cite{nutshell} for a similar construction in 3d.

The third ingredient we need is the graviton field operator.  This is the fluctuation of the metric operator over the flat metric $h^{ab}(\vec{x}) = g^{ab}(\vec{x}) - \delta^{ab}$. It is more convenient to consider here the fluctuation of the densitized metric operator $\tilde h^{ab}(\vec{x}) = (\det{g}) g^{ab}(\vec{x}) - \delta^{ab}= E^{ai}(\vec{x})E^{bi}(\vec{x}) - \delta^{ab}$.  In the linear theory, the
propagators of the two agree because of the trace-free condition.  To determine its action, we can equally use the geometrical interpretation discussed above, or LQG. We identify the point $\vec{x}$ with one of the nodes $n$ of the boundary spinnetwork $s$.  Equivalently, with (the center of) one of the tetrahedra of the triangulation.  Four links emerge from this node.  Say these are $e_I, I=1,2,3,4$. They are dual to the triangular faces of the corresponding tetrahedron. Let $n^I_a$ be the oriented normal to this face, defined as the vector product of two sides. Then $E(n)^{Ii}=E^{ai}(\vec{x})n_a^I$ can be identified with the action of the $SU(2)$ generator $J^i$ on the link $e_I$. We have then immediately that the diagonal terms define diagonal operators \cite{lqg,book}
\begin{equation}
E^{Ii}(n)E^I_{i}(n)|s\rangle = (8\pi\hbar G)^2\ j_{I} (j_{I} +1)|s\rangle ,
\label{diagonal}
\end{equation}
where $j_I$ is the spin of the link $I$.  Non--diagonal terms, that act on the intertwiners, are discussed elsewhere.

We have now all the elements needed to compute the expression (\ref{simplypropaga2}).  Inserting 
(\ref{simply1}), (\ref{vuoto}) and (\ref{diagonal}) into  (\ref{simplypropaga2}) we obtain a well--defined expression for the propagator.   We choose the points $x$ and $y$ to be two distinct nodes of the boundary spinnetwork.  Equivalently, these can be thought as points located, say, in the centers of the corresponding dual tetrahedra: in the theory, of course, position is not determined with precision lower that the individual ``atoms  of spaces" described by the individual tetrahedra.  We consider the ten by ten matrix $\widetilde W(L)$ formed by the ``diagonal" components of the propagator  
\begin{equation}
\widetilde W(L)_{(ij)(kl)}\equiv  W^{abcd}(x,y;q)\ n_a n_b\  \tilde n_c \tilde n_d,
\end{equation}
where $n=n_{ij}$ and $\tilde n=n_{kl}$, and $n_{ij}$ is the normal to the triangle $t_{ij}$.  
We also write ${W(L)}  \equiv \widetilde W(L)/{|n|^4}$. By symmetry
\begin{equation}
W(L)_{(ij)(kl)}= W_1(L)\ a_{(ij)(kl)}- W_2(L)\ \delta_{(ij)(kl)}  +  W_3(L)\ b_{(ij)(kl)}. 
\label{simmetry2}
\end{equation}

Before computing these quantity in the background independent theory, let us compute it in conventional linearized quantum general relativity.  
In a flat background metric, two points in the center of adjacent tetrahedra, in a surface with the boundary geometry chosen, are at a distance ${|x^u_1-x^d_1|_q}=L/4$. 
If the four indices $i,j,k,l$ are all distinct, it is easy to see that $n$ and $\tilde n$ are orthogonal; 
then the propagator is easily computed to be \cite{Mattei:2005cm} 
\begin{equation}
W_{(ij)(kl)}^{\rm linearized}(L) =i\frac {8\pi\hbar G}{4\pi^2}\frac{1}{|x^u_1-x^d_1|^2_q}=i\frac{32 \hbar G}{\pi L^2} 
\label{sivorrebbe1}
\end{equation}
On the other hand, the components $W^{\rm linearized}_{(ij)(ij)}$ and $W^{\rm linearized}_{(ij)(ik)}$ are vacuum expectation values at fixed ``time": the first is the flucuation of the area square of a triangle, and the second is the vacuum correlation between the fluctuations of the area squares of two adjacent triangles in the same tetrahedron.  These are also proportional to $L^{-2}$. We can therefore write
\begin{equation}
W^{\rm linearized}(L) =\frac{32 \hbar G}{\pi L^2}\ {\cal W} 
\label{sivorrebbe}
\end{equation}
where ${\cal W}$ is a numerical matrix, with the same symmetry structure as in (\ref{simmetry2}). In particular, ${\cal W}_{(12)(34)}={\cal W}_3=i$, while ${\cal W}_1$ and ${\cal W}_2$ are real  numbers of the order of unity, easily obtained from the linear theory. 

We now compute the matrix $W(L)$  in the full theory. Since this is a diagonal term in the propagator, we can use (\ref{diagonal}) and (\ref{simplypropaga2}) reads
\begin{equation}
\widetilde W(L)_{(ij)(kl)}=\sum_s W[s] \ 
((8\pi\hbar G)^2 j_{ij}(j_{ij}+1)-|n|^2)
((8\pi\hbar G)^2 j_{kl}(j_{kl}+1)-|\tilde n|^2)
\Psi_q[s].
\end{equation}
The terms $|n|^2$ come from the background $\delta^{ab}$ and are equal to the square of the area of the face, namely to $(8\pi\hbar G j_{L})^2$, for large $j$.   Inserting (\ref{simply1}) and (\ref{vuoto}) we have 
\begin{eqnarray}
\widetilde W(L)_{(ij)(kl)}&={\cal N}  
\frac{\lambda(8\pi\hbar G)^4}{5!}& \sum_{j_{nm}} 
 \left(\prod_{n < m} {\rm dim}({j_{nm}})\right) \, 
(j_{ij}(j_{ij}+1)-j_{L}^2) \  ( j_{kl}(j_{kl}+1)-j_{L}^2) \ \ 
                         \nonumber\\   && 
                          \mathcal{B}(j_{nm})\ 
                         \ \exp \left\{
-\frac{1}{2j_L}\alpha_{(nm)(pq)}(j_{nm}- j_{L})(j_{pq}- j_{L})
+i\Phi \sum_{n,m} j_{nm}\right\}. 
\label{full}
\end{eqnarray} 
Since the vacuum exponential peaks the sum around $j_{L}$, which is large, we can discard the $+1$. We expand the summand in the fluctuations $\delta j_{ij}=(j_{ij}-j_{L})$, and keep only the lowest term, assuming that the gaussian suppress the higher terms.  We redefine $\cal N$ absorbing all the factors that cancel with terms in (the expansion to this order of) the partition function.  We assume that the ${\rm dim} j$ terms vary slowly over the range where the gaussian is peaked, and can be considered constant. We change summation variable from the spins to the fluctuation of the spins 
\begin{equation} 
\widetilde W(L)_{(ij)(kl)}=4{\cal N}(8\pi\hbar G)^4j_L^{2}  
\sum_{\delta j_{nm}} 
\delta j_{ij}\ \ \delta j_{kl} 
                     \    \mathcal{B}(j_L+\delta j_{nm})\ 
                         e^{-\frac{1}{2j_L}\alpha_{(nm)(pq)}\delta j_{nm}\delta j_{pq}
                                                  +i\Phi \sum_{nm}j_{nm}}. 
\label{somma}
\end{equation} 
The rapidly oscillating term $\exp {i\Phi \sum_{nm}j_{nm}}$ tends to suppress the sum.  To evaluate it, we need the explicit form of $\mathcal{B}(j_L+\delta j_{nm})$ in the large $j$ regime. 
This is of the form \cite{BCE,FL} 
\begin{equation}
B(j_{nm})
= \sum_\tau P_{\tau} \left[e^{iS_{\rm Regge}(\tau)+k_\tau\frac{\pi}{4}}+e^{-iS_{\rm Regge}(\tau)+k_\tau\frac{\pi}{4}}\right]+D(j_{nm}) 
\label{ansatz}
\end{equation}
Here $\tau$ labels the distinct  4-simplices having areas $A_{nm}=\sqrt{j_{nm}(j_{nm}+1)}$ and  $P_{\tau}$ is a slowly varying factor. Since the sum (\ref{somma}) is peaked around $j_{nm}=j_L$, let us expand the 10$j$ symbol around this point. To second order around $j_{nm}=j_L$, the Regge action reads 
\begin{equation}
S_{\rm Regge}(j_{nm}) = \Phi \sum_{nm} j_{nm} + \frac12 G_{(mn)(pq)} \delta j_{mn}\delta j_{pq}, 
\end{equation}
where, introducing the ``discrete derivative"
$\frac{\partial f(j)}{\partial j}\equiv \
f(j+1/2)-f(j)$, we have defined
\begin{equation}
G_{(mn)(pq)}= \left.\frac{\partial\Phi_{mn}(j_{rs})}{\partial j_{pq}}\right|_{j_{rs}=j_L}. 
\label{angolispin}
\end{equation}
Thus, around  $j_{nm}=j_L$, we have 
\begin{equation}
B(j_{nm})
= P_{\tau_R} \left[e^{i(\Phi\! \sum_{nm} j_{nm} + \frac12 G_{(nm)(pq)} \delta j_{nm}\delta j_{pq}+\frac{\pi}{4})}\!+e^{-i( \Phi\! \sum_{nm} j_{nm} + \frac12 G_{(nm)(pq)} \delta j_{nm}\delta j_{pq}+\frac{\pi}{4})}\right]\!+
D(j_{nm})
\end{equation}
where $\tau_R$ is the regular four simplex (for which $k_{\tau_R}=1$), which is only non-degenerate four-simplex with these areas \cite{BCE}.  The key observation is now the fact that the rapidly oscillating $\exp \{i\Phi \sum_{nm}j_{nm}\}$ term in the second term of this expression cancels with the rapidly oscillating term in (\ref{somma}).  Therefore the second term of the last expression contributes in a non-negligible way to the sum 
(\ref{somma}). The first term is suppressed (by the rapidly oscillating factor $\exp {2i\Phi \sum_{nm}j_{nm}}$) and it is reasonable to expect that so is the degenerate term $D(j_{nm})$
because this corresponds to 4-simplices with different angles, and should
be dominated by different frequencies.   Therefore  (\ref{somma}) becomes, keeping 
only the first term 
\begin{equation} 
W(L)_{(ij)(kl)}=4{\cal N}j_L^{-2} P(\tau_R) e^{\frac{i\pi}{4}}
\sum_{\delta j_{nm}} 
\delta j_{ij}\ \ \delta j_{kl}  
\ e^{\frac{i}{2} G_{(nm)(pq)} \delta j_{nm}\delta j_{pq}}
                         e^{-\frac{1}{2j_L}\alpha_{(nm)(pq)}\delta j_{nm}\delta j_{pq}},
\end{equation} 
where we have also used $\widetilde W(L)=|n|^4 W(L)=(8\pi\hbar Gj_L)^4 W(L)$. Approximating the sum by a gaussian integral gives
\begin{equation} 
W(L)
=4j_L^{-2}  \left(j^{-1}_L\alpha - iG\right)^{-1}
\label{eccola}
\end{equation} 
We only need to evaluate the derivatives (\ref{angolispin}) of the angles with respect to the spins.   4d geometry gives  \cite{areaangle}
\begin{equation}
\frac{\partial \Phi_{nm}}{\partial A_{pq}}= \frac{1}{\sqrt{5}L^2}\left(\frac72 a_{(nm)(pq)}-9 \delta_{(nm)(pq)}  -4 b_{(nm)(pq)}\right)\equiv \frac{1}{L^2}\ {\cal K}_{(nm)(pq)}.
\end{equation}
The ten by ten matrix ${\cal K}$ has purely numerical entries. From the relation between areas and spins, we have
\begin{equation}
G_{(nm)(pq)}= {8\pi\hbar G}L^{-2}{\cal K}_{(nm)(pq)}
= \sqrt{3}(4j_L)^{-1}{\cal K}_{(nm)(pq)}.
\end{equation}
Notice the $j_L$ factor that combines with the one in front of $\alpha$ in 
(\ref{eccola}) to give the crucial overall $1/j_L$ dependence of the propagator.
Using the reation between spins and areas, (\ref{eccola}) reads
\begin{equation} 
W(L)
=\frac{32\pi\hbar G}{\sqrt{3}/4\; L^2}\left(\alpha+ i\sqrt{3}/4\; {\cal K}\right)^{-1}. 
\label{eccola2}
\end{equation} 
This is precisely the value  (\ref{sivorrebbe1}-\ref{sivorrebbe}) of the propagator computed from the linearized theory, with the correct $1/|x-y|^2$ spacetime dependence. The three numerical coefficients of the matrix $\alpha$ are completely determined by $\alpha={{4\pi^2}}/{\sqrt{3}} \ {\cal W}^{-1} -i{\sqrt{3}}/{4}\ {\cal K}$. 

The propagator we have computed approximates the perturbative one. This is correct for large $|x-y|$, where the approximations used in evaluating the sum (\ref{full}) hold.  When $|x-y|$ approaches the Planck length, quantum gravitational corrections appear. In this regime, (\ref{full}) can be easily computed numerically.  Given the discreteness of the representation, the propagator is likely to be cut-off at the Planck scale.  Therefore we obtain the result that, at least to this order, the (component we have computed of the) propagator is the standard free-theory one at large distances, with corrections at the Planck scale, wilch appear to make it finite.

Our main results are: (i)  the ``bad" terms of the Barrett-Crane vertex amplitude are cancelled when the correct vacuum state, appropriately peaked on the extrinsic geometry, is used; (ii)  the correct propagator emerges to this order; and (iii) --for us the most interesting result-- it appears to be possible to compute $n$-point functions from background--independent quantum field theories.

\centerline{------}

I am strongly indebted to Leonardo Modesto, Simone Speziale, John Baez and Dan Christensen for crucial criticisms, inputs and suggestions.

\end{document}